# Knowledge Utilization and Open Science Policies:
# Noble aims that ensure quality research
# *or*
# Ordering discoveries like a pizza?

## Julia Heuritsch[a]

[a] German Centre of Higher Education Research and Science Studies (DZHW), Schützenstraße 6A, 10117 Berlin;
Humboldt Universität zu Berlin, Research Group "Reflexive Metrics", Institut für Sozialwissenschaften,
Unter den Linden 6, 10099 Berlin, heuritsch@dzhw.eu

### Abstract

Open Science has been a rising theme in the landscape of science policy in recent years. The goal is to make research that emerges from publicly funded science to become findable, accessible, interoperable and reusable (FAIR) for use by other researchers. Knowledge utilization policies aim to efficiently make scientific knowledge beneficial for society at large. This paper demonstrates how Astronomy aspires to be open and transparent given their criteria for high research quality, which aim at pushing knowledge forward and clear communication of findings. However, the use of quantitative metrics in research evaluation puts pressure on the researcher, such that taking the extra time for transparent publishing of data and results is difficult, given that astronomers aren't rewarded for the quality of research papers, but rather their quantity. This paper explores the current mode of openness in Astronomy and how incentives due to funding, publication practices and indicators affect this field. The paper concludes with some recommendations on how policies such as making science more "open" have the potential to contribute to scientific quality in Astronomy.

## 1. Introduction

Making science more open has been a rising theme for policy stakeholders at international (e.g. European Commission [1] and United Nations [2]) and national levels (e.g. Netherlands Organisation for Scientific Research; NWO [3]). The goal is to make research that emerges from publicly funded science to become findable, accessible, interoperable and reusable (FAIR) for use by other researchers. Stakeholders like the NWO and European Commission acknowledge that the way science is conducted is fundamentally changing due to the sophisticated digital technologies available. Often these 'open' policies are part of a more general 'knowledge utilization' policy, which aims to efficiently make scientific knowledge beneficial for society at large under the imperative of the Three Os – Open Innovation, Open Science and Open to the World. The aim of the paper is to give recommendations of what 'open' policies would need to consider in order to support FAIR publication in Astronomy, while at the same time encouraging quality in the knowledge production process. This investigation includes in what way astronomers value openness and how current policies encourage openness and research quality.

New policies may come with new incentives and new ways to measure whether certain goals have been achieved. These may not only come with the aspired effects, but also with "unintended consequences" (drawing back to the notion of "unanticipated consequences of purposive social action" [1]) such as quality discrimination. The notion of unintended consequences has been a long-standing debate in the sociology of (e-)valuation as well as in the contexts of notions of "reflexivity", "performativity", self-fulfilling prophecies or "retroaction". All of these approaches share a common denominator, namely that there seems to be a disjunctive moment between intention, action not only on the micro-level of the individual actor, but also on the level of agglomerations spanning from communities of practice up to society at large. It might therefore be valuable to approach the issue of discourse on a certain topic, in this case "open science" with the notion of what in the most simple formula be termed "feedback" and at the same time the relation between justification and critique that structures part of that discourse.

One aspect of justification and critique becomes apparent in questions how to measure and ensure high-quality of knowledge production in the context of "*Accountability*" and "*transparency*". Both concepts are closely associated with producing and monitoring metrics [2]. This is because quantification is one means to constitute social entities as things that last and are comparable. As such, the goal of quantification is to enable objectification and to master uncertainty. Through objectification, both a political space and a measuring space, are co-constituted in which things can be compared [3]. It permits scrutiny of complex or disparate phenomena in ways that enable judgment [4]. Hence, quantification offers a shared language and replaces trust in people with "trust in numbers" [5]. Quantification therefore is seen as one mode of de-localization of valuation practices. Quantitative metrics, such as indicators to measure scientific productivity, *commensurate*, which is the act of using numbers to rate and rank, creating a specific type of relationship among objects. Commensuration is one of the most consequential uses of numbers [4], because it turns describing numbers into prescriptive ones. Commensuration attributes meaning to numbers.

*Effort* or how individual researchers *perform* cannot be monitored efficiently so assessment cannot be based upon it and therefore, a scientist is rewarded and funded for quantitative achievement instead [6]. The same holds for *usefulness* of science. Not only is it difficult to measure usefulness, but there is also a tension between the ever increasing demand for *societal relevance* (e.g. [7]) and the risky nature of basic research. The problem of commensurability of usefulness also becomes apparent, when we observe typical means of addressing usefulness or impact in research evaluation. One of the most notable examples being the narrative form of signifying impact in the Research Excellence Framework (REF). Yet, as elaborated before, there are social forces that adhere to the idea of measurability. Basic research is the human's endeavour to understand the unknown and as such it is by definition risky [8]. When societal relevance is measured in applicable outputs, and decisions must be made about distribution of researchers amongst research fields, economic pressure to produce such outputs can arise. This may lead to a tension between demonstrating its usefulness to society, on the one hand, and not being able to guarantee that due to the risky nature of research, on the other hand. For research fields that perform mainly basic research, such as Physics and Astronomy, this pressure to justify their societal relevance might be especially high, since results of basic research may have a delay in leading to ground breaking technological developments or theories. More applied sciences, "where the products of research are highly profitable, such as medicine, biotechnology, genetics and military research" [9], face a different economic pressure, the expectation to produce more and better.

Measures that initially may have been designed to describe behaviour can easily be used to judge and control it, due their commensurative character. Measurement intervenes in the social worlds it depicts, as measures are *reactive & performative*; they cause people to think and act differently. Hence, numbers can also exert discipline on those they depict and disciplinary practices define what is appropriate, normal, and to what we should aspire [4]. Foucault ([10] & [11]) links statistical practices to "*governmentality*", a term to describe how the government uses numbers to influence citizens so that they fulfil those government's policies. He describes discipline as a mode of modern power that is continuous, diffuse and embedded in everyday routines.

The performative character of indicators leads to the conclusion that policies have constitutive effects on how science is done [12]. In other words, indicators and rewards introduced by policies can shape the process how knowledge is produced in science. The fact that indicators commensurate, where all difference is transformed into quantity [4], leads to the argument that their use to assess scientific quality gives rise to an "evaluation gap". This is a term coined by Wouters [13] to acknowledge a discrepancy between the notions of scientific quality as perceived by researchers of a field and as measured by indicators. In order to meet the targets set by indicators, scientific quality may be sacrificed. This can have "unintended consequences" such as goal displacement, gaming, information overload, questionable authorship practices, unhealthy competition and aversion to risky/innovative projects ([14], [15]).

As outlined above, basic research is under the pressure of demonstrating its societal relevance. While astronomy asks highly fundamental questions, which inspire both scientists and the public at large, Astronomy faces a crisis to demonstrate its usefulness. That is shown by the fact that one can find a large number of flyers and commentaries on the internet and elsewhere explaining why astronomy is important (e.g. [16]). Applications from Astronomy are rather invisible for the public, due to two reasons. First, as outlined above, basic research usually doesn't find immediate societal usefulness. An example is applying the theory of general relativity to enable precision in the Global Positioning System (GPS). Second, uncovering the laws of nature, such as the theory of general relativity, is Astronomy's primary goal, resulting applications are secondary. Nevertheless, technology invented for space exploration often leads to surprising applications on Earth, so-called spin-offs, such as innovations in dental care and breast cancer detection. Hence, there is a tension between needing an output due to economic pressures and doing basic research for its own sake that entails dead ends and no immediate usefulness. Additionally, large and expensive (international) observing instruments involve large collaborations and the use of (open) archives and huge datasets. For these reasons, Astronomy is such an interesting case to study the significance of indicators in its usefulness crisis and their relevance in this tension relationship. To which extent indicators shape the conditions under which astronomers produce knowledge, is what needs to be answered when developing new policies.

**2. Methods**

This research[4] consists of semi-structured interviews and a document analysis. The interview sample was targeted such that the interviewees represent a variety of nations, career-status and research areas. 4 faculty members, 2 postdocs, 1 PhD and 2 Master students from Leiden Observatory (Sterrewacht) were chosen who work in cosmology (2), exoplanets (1) and observational (radio) astronomy in different subfields (6). The Master programme at the Sterrewacht is very research intensive, requiring the student to write two Master theses in total, which is the reason why they are also interesting subjects for this study. In order to shed light on what effects policies have on the field, questions were developed such that an astronomer's definition of quality versus what is measured by indicators can be studied. Next to the meaning of openness and data sharing, topics include career steps, project funding, exposure to assessments, research evaluation, the publication and funding system, different stages of the knowledge production process – from planning, via doing the research to publishing – and the meaning of quality. Each topic was introduced by one overarching question, followed by several potential follow-up questions.

The participating researchers were invited via email and all names are anonymized. All interviews, 80-100 minutes in length, were fully transcribed into electronic form, summarised and coded) according to Grounded Theory. These codes represent themes which emerged by combining sensitivity towards existing literature on constitutive effects of indicator use with insights from the data.

The interview data were complemented with a document analysis of materials collected online or made available via the informants, including CVs of the interviewed researchers, annual reports (1998 to 2015), (self-) evaluation reports of the Dutch Astronomy institutes and their umbrella organisation NOVA. The documents informed the following analysis, however the document analysis itself is not part of this study and can be found in the original report of the project[4].

In the *Results* section direct quotes of the interviewees will be given between double quotation marks.

## 3. Results
*3.1 The astronomers' stance to openness and effects of current policies*

In order to understand what policies related to 'opennes' would mean for Astronomy, we must first ask what data and research results mean to an astronomer. My study found that astronomers generally conduct science for the sake of curiosity and "pushing knowledge forward" (Faculty Member 4). For them, publications are not their priority, but rather a means to publicise their results in order to advance the field in three steps:

> "You have a new science idea. You have asked the question clearly and well, with a well-defined […]. And you have written a paper which demonstrates you have answered that question […]. And you have written it in such way that a non-expert in that field can read it and understand what you have done. […] If it's a crap written paper, then that's crap research – I don't care how brilliant the answer is, if they can't communicate it through a paper or through a presentation, then that's bad research." (Faculty Member 4)

> "And ahh … if that was not so important [to get papers out] I would probably not bother so much … I mean I would still publish my papers because I – it gives a different motivation to it, right? As a scientist you just want to publish your papers, because you are a scientist and you think this is important for science: 'This is the result, this is what defines the process of science'." (Faculty Member 1)

Rather, it is in the astronomers' interest to share data and results to get knowledge "out to the community" (Faculty Member 1). "To know and understand better" (Postdoc 1) and communicate this knowledge to the community is what makes up an astronomer's intrinsic motivation to conduct research. For an astronomer, the definition of high-quality research is based on this motivation. I found three quality criteria:

- Asking an important question for the sake of understanding the universe better and to push knowledge forward.
- Using clear, verifiable and sound methodology.
- Clear communication of the results so the community can make use of them.

From these criteria it is apparent that astronomers' motivation in doing science is not for some direct societal impact, since they are dedicated to fundamental research which might only find applications decades later [8].

> "Well, academic quality has always been relatively clear. It has to be verifiable and clear, unbiased etc. But there is these days … a tendency to look at the value of science in terms of economic output, it's called '*valorization*'. And I am totally uninterested in that […]. It is always nice if you find applications that are useful […]. Why not? But that's not why we do it." (Faculty Member 1)

On the other hand, openness in terms of disseminating the knowledge gained does fit within astronomers' values and definition of quality. To understand how 'open' policies could have positive effects on openness in Astronomy, and whether they could at the same time encourage scientific quality, we must first look at the effects of current policies.

This study found factors for extrinsic motivation that drive research in Astronomy as well as the intrinsic factors. Extrinsic motivation arises from what the evaluation system values through its indicators. Astronomers report that *first author publications*, *citation rates* and *number of acquired grants* are what determine their value as a researcher. Since the future career of an astronomer is dependent on these factors, there is a shift from the initial motivation to publish for the sake of disseminating knowledge, to the "need to publish" (e.g. Postdoc 1 & Faculty Member 1). This results in publication pressure and lower quality papers.

> "Your job prospects will depend on this like quantity rate, with which you are publishing." (PhD Candidate)

> "It's a system problem I think. Erm, I try to do quality research, but I do feel sometimes that I end up publishing because I have to publish.", "I wish we could just focus on more like quality papers instead of quantity papers." (Postdoc 1)

Observational astronomers are found to be particularly affected by this pressure as compared to theoretical astronomers. On the one hand, they produce data with telescopes which are essential for the knowledge production process in Astronomy. On the other hand, this data is a form of output that is reportedly not valued in evaluations. To produce data, observational astronomers need to compete for limited *observation time* at telescopes and then 'be lucky' to have the right weather conditions. Once granted, observation time does count like received funding in an astronomer's CV. However, non-detections are more common than detections and 90% of non-detections are not publishable. Hence, observational astronomers face the risk to fall through the cracks of metrics in every step of their knowledge production process.

> "So it's essentially [that] negative results are considered as failed research by the community. […] And on that side I disagree. […] So there is always information to be taken from research that is well conducted. Given that the research is using state of the art data, and state of the art methodology, whatever the result is, should be interesting." (Postdoc 2)

"Non-detections. [...] It's just really hard to work with the telescope and I really want to be able to figure it out and do this thing and I think personally I would feel failed if I wouldn't be able to at least … put some limit, that gives a good sort of low sensitivity to it [i.e. finding some implications]." (PhD Candidate)

*3.2 Current state of openness in Astronomy*

The effects of current policies affect not only research quality, but also how *FAIR* astronomers' research data and results are. To demonstrate the current state of openness in Astronomy, output can be divided into three categories: *raw data*, *results* and *negative results*. Policies which advocate for openness and good research quality in Astronomy would have to take these as a starting point.

First, *raw*, unprocessed data is the data to be analysed by the researcher. The most prominent example is telescope data. Most telescopes make the raw telescope data public after a proprietary period to the original observer of one year after obtaining through the archives of the observatories. Additionally, researchers can publish their raw or reduced data through the archive of Centre Donnees Stellaire in Strasbourg or various others, once their paper is accepted. On the one hand, this serves the purpose of openness and gives other researchers the possibility to replicate or conduct their own studies. On the other hand, the relative short proprietary period of one year adds to the publication pressure, as only the first to publish receives the credit.

Second, *results* are written up in publications in the form of scientific articles. In a first step of the data analysis, raw data is processed to reduced data through so-called 'pipelines', which are data reduction codes that clean the data from noise and prepare them for the analysis. Sometimes these reduced data are also published in the publically available observatory archives. Results are the final processed data and the conclusions drawn from them. In order to have impact, astronomers publish in journals, but they usually also upload their papers to the open data repository ArXiv. However, *findable* results does not equal *accessible*, even for fellow researchers. The reasons for non-accessibility are for example that reduction codes are mostly not published or that important steps in calculations are not mentioned in the paper, decreasing transparency and hence the paper's communication value. Interviewees urge for more reader-friendly formats, which would enable, for example, expanding sections, interlinking content, adding simulations/ visualisations and publishing code, as it would be possible with modern technology. Despite this, papers are still written in a style inherited from pre-computer times. As information about methods and analysis gets lost with result oriented papers, written in an out-dated manner, it is more difficult to ensure good communication and replicability of research results. In some cases, mistakes even remain undiscovered.

> "The way that papers are currently being written is perhaps too much tied to the way that papers were published in the past. So they were actual papers in a journal, so they had to be sequential. But this is no more the case, now that we have other ways to … read or get information. We can have, not necessarily interactive things, but, at least content that can be separated into different sources. So you can read on one side about the science of the paper, and on the other side about the technical aspects. And currently the two things are merged into a single file, or work. And even if it's true that you intent to have sections like methodologies and results, so if you are not interested in the methodology, or if you actually want to read about the methodology you can go there or not go there. But people will tend to get take [content] away from the methodology section, because they will consider 'Ah, that's too much […], so let's not mention this or put that into an appendix'. So I think there should be the possibility for authors to be very thorough in explaining the methods and even, that includes the possibility to show code. […] In fact in the Astrophysics community, the skills in programming are fairly low in general. Which is worrysome, because I think there are a lot of bugs running around that are not noticed. And because we can't look at the code we can't say, or see whether this is happening or not." (Postdoc 2)

As mentioned above, raw or reduced data is often published along in various archives. However this process is not standardized and often voluntary. While astronomers cite their (data) sources, interviewees criticise that references are not transitive. For example, a literature review might cite the papers it is referring too, but not the data sources that these papers are based on. This seems hardly *fair* for the producers of the data and doesn't add to transparency of the research process. Since there are no incentives for replicability or for transparency in one's research methods, and since publication pressure is high, astronomers do not have the time to consider whether their results are fully *reusable*:

> "And I think that's very bad, when for example, almost the entire results come from a code which is not publically available. So you cannot look at this code and see … if they are actually doing what they say in the paper. And also if they – sometimes they make a mistake. […] So in the sense, the replicability of the work we do … is not always very high. And in the sense that you can download for yourself, in principle all the raw datasets from a telescope and you can redo everything by yourself. So in this sense, yes it's replicable, but never fully replicable." (Postdoc 2)

Third, conducting fundamental research can naturally lead to a dead end, or to *negative results*. The most common examples in Astronomy are non-detections, which are observations that didn't lead to the predicted detection. They are usually only publishable when the researcher can determine their implication or is able to provide upper or lower limits.

> "[Non-detections are not publishable], unless you have a very good [implication], as in for example the way we sort of explained the upper limits with the non-detection. […] The problem is how to tailor it, right? […] So, yeah, unless you have … like a good way, I mean there is some research that published non-detection – for exoplanets sometimes they

> publish it when they didn't detect it, because sometimes you sort of predict that it should be there […] And it's an anomaly or something like that […] So there are some ways to publish this, but I think it's very … like 10%. There is a whole 90% that doesn't get published and sometimes, like for example, if you just had bad weather, then it's very difficult, right?" (Postdoc 1)

In some sub-fields of Astronomy, non-detections are more common than detections. Most interviewees are convinced that negative results should "absolutely be publishable" (Faculty Member 4). This is because they are seen as valuable with respect to new knowledge about what does not work. As research is the discovery of the unknown, this kind of information is also essential, "because it either can help [the researchers] discount certain theories, or help them kind of support other theories" (Master Student 2). Hence, astronomers are advocating for the exposure of the negative results as well, so that other researchers don't have to 'reinvent the wheel'.

> "I mean when I was at [the famous institution] we said, we should start a journal on non-detections. Because I am really sure that there are people that have been observing the same objects on and on and in without knowing that other people have already done this. Because nobody published when they don't detect anything." (Postdoc 1)

While efforts like this would be welcomed by astronomers, current metrics do not account for non-detections and hence there are little incentives to invest time in the contribution to such 'non-detection journals' if there are no benefits for one's further career in science.

### 4. Conclusion & Recommendation

The results of this study show that Astronomy aspires to be open and transparent. However, this requires support from policy makers as current policies do not provide the incentives to invest valuable time in the publication of data and code. Therefore, the observations above imply five recommendations to policy makers when it comes to knowledge utilization and openness:

1. Goodhart's law which states 'when a measure becomes a target, it ceases to be a good measure', always needs to be kept in mind when establishing new policies. In practice this means, that indicators and rewards have constitutive effects on knowledge production, which need to be accounted for.

2. For the afore-mentioned reasons, *FAIR*ness cannot be implemented if there is no incentive for a change in the way research is published. This may require a change in journal templates and paper-writing style to take advantage of the possibilities offered by modern technology. As one of the astronomer's quality criteria are to communicate results well and transparently, papers would be of higher quality if they included more information on in-between-steps; this could be done through expandable sections that a reader could easily skip if wanted. Incentives for including code would lead to a decrease the propagation of errors and increase *replicability*, *accessibility* & *reusability*. If publications provided an interactive way of delivering visible feedback and updating outdated information, *reusability* would benefit and an active exchange within the scientific community would be fostered.

3. Journals could support the astronomers' aspiration for knowledge utilization in terms of pushing knowledge forward by building on previous research. Providing for transitive referencing would make the sources that pieces of research is based on more transparent. 'Open' policies may encourage the journals to do so and the astronomers to use these opportunities.

4. Advocating for openness and knowledge utilization also means valuing knowledge about what doesn't work. Therefore, policies need to reward the publication of research that led to negative results to ensure that researchers who engaged in those studies receive credit for their work, thereby reducing publication pressure. Especially for observational astronomers in sub-fields where the majority of the detections currently are not publishable, 'open' policies provide a hope for improvement.

5. Unlike Astronomy, applied sciences may naturally be more orientated towards finding economic applications for their research. However, this quote applies for all sciences, and is to be kept in mind when it comes to policies for knowledge utilization:

"[These politicians] think that they can direct science. […] – *They think they can order discoveries like you order a pizza*. You. Cannot. Order. A. Discovery. […] You have to work on it, you have to try things, you have to experiment. […] But since science is funded mostly by public funding, we are dependent on the strange conceptions that politicians have on how science works." (Faculty Member 3)

**Acknowledgements**

I would like to thank the interviewees for their willingness to participate in this study and for dedicating research time that is even more valuable in times of publication pressure.

**References**

[1] Merton, R. (1936), "The unanticipated consequences of purposive social action", *American Sociological Review,* Vol. 1, No. 6 (Dec., 1936), p. 894-904

[2] Espeland, W.N. & Vannebo B. (2008), "Accountability, Quantification, and Law", *Annual Review of Law and Social Science 3*, p. 21-43

[3] Desrosières, A. (1998), "The Politics of Large Numbers – A History of Statistical Reasoning", *Harvard University Press*, ISBN 9780674009691

[4] Espeland, W.N. & Stevens, M.L. (2008), "A Sociology of Quantification"**,** *European Journal of Sociology*, Volume 49, Issue 03, p. 401 – 436, DOI: 10.1017/S0003975609000150

[5] Porter, T. (1995), "Trust in numbers", *Princeton University Press*

[6] Rosenberg, N. & Nelson, R. (1994), "American Universities and technical advance in

industry", Research Policy 32, p.323-348

[7] Lex M. Bouter (2008), Knowledge as Public Property: The Societal Relevance of Scientific Research, OECD, http://www.oecd.org/site/eduimhe08/41203349.pdf

[8] Stephan, P. (2012), "How economics shapes science", Harvard University Press

[9] Bourdieu, P. (2004), "Science of Science and Reflexivity", The University of Chicago Press, ISBN: 9780226067377 & ISBN: 9780226067384

[10] Foucault, M. (1977), "Discipline and Punish: The Birth of the Prison", London, Allen Lane

[11] Foucault, M. (2003), "The Subject and Power", in Rabinow Paul and Nicholas Rose, eds., "The Essential Foucault" (New York, The New Press, p. 129-144)

[12] Dahler-Larsen, P. (2014), "Constitutive Effects of Performance Indicators: Getting beyond unintended consequences", Public Management Review, 16:7, p.969-986

[13] Wouters, P. (2017), "Bridging the Evaluation Gap", Engaging Science, Technology, and Society 3: p.108-118

[14] Rushforth, A.D. & De Rijcke, S. (2015). "Accounting for Impact? The Journal Impact Factor and the Making of Biomedical Research in the Netherlands", Minerva 53, p.117-139

[15] Laudel, G. & Gläser, J. (2014), "Beyond breakthrough research: Epistemic properties of research and their consequences for research funding", Research Policy 43, p.1204-1216

[16] Rosenberg, M. et al. (2013), "Why is astronomy important?", https://arxiv.org/abs/1311.0508

---

[1] https://ec.europa.eu/research/openscience/index.cfm
[2] http://www.unoosa.org/oosa/en/ourwork/psa/schedule/2017/workshop_italy_openuniverse.html
[3] https://www.nwo.nl/en/policies/knowledge+utilisation
[4] The complete report of the results of this study can be found on ArXiv (https://arxiv.org/abs/1801.08033), however this paper concentrates on what the findings mean for the latest policy developments regarding the *FAIR* imperative.